# Scaling relation for earthquake networks

S. ABE[1,2] and N. SUZUKI[3]

[1] *Department of Physical Engineering, Mie University, Mie 514-8507, Japan*

[2] *Institut Supérieur des Matériaux et Mécaniques Avancés, 44 F. A. Bartholdi, 72000 Le Mans, France*

[3] *College of Science and Technology, Nihon University, Chiba 274-8501, Japan*

**Abstract.** — The scaling relation derived by Dorogovtsev, Goltsev, Mendes and Samukhin [*Phys. Rev. E*, 68 (2003) 046109] states that the exponents of the power-law connectivity distribution, $\gamma$, and the power-law eigenvalue distribution of the adjacency matrix, $\delta$, of a locally treelike scale-free network satisfy $2\gamma - \delta = 1$ in the mean field approximation. Here, it is shown that this relation holds well for the reduced simple earthquake networks (without tadpole-loops and multiple edges) constructed from the seismic data taken from California and Japan. The result is interpreted from the viewpoint of the hierarchical organization of the earthquake networks.

In a real complex system, detailed information on properties of system elements and their interaction/correlation may not always be available. In such a situation, the network description offers a useful tool. There, vertices and edges connecting them effectively represent elements and their interaction/correlation, respectively. Analyzing the structure of such a network and dynamics on it, one is able to grasp the gross property of the system.

Recently, the concept of complex networks has been introduced to the field of seismology [1]. The construction of an earthquake network proposed there is simple. A geographical region under consideration is divided into a lot of small cubic cells of certain size. A cell is regarded as a vertex if earthquakes with any values of magnitude occurred therein. Two successive events define an edge between two vertices. This edge effectively replaces complex event-event correlation, the mechanism of which is largely unknown. If two successive earthquakes occur in the same cell, they form a tadpole-loop (or, a bubble). This procedure enables one to map seismic data to a growing stochastic network.

Some comments on the above-mentioned construction are in order. First of all, it contains a single parameter: the cell size, which is the scale of coarse graining. Once the cell size is fixed, an earthquake network is unambiguously constructed from seismic data. However, since there exist no *a priori* operational rules to determine the cell size, it is of importance to examine how the property of an earthquake network depends on this parameter. Secondly, a full earthquake network is a directed one in its nature. Directedness does not bring any difficulties to analysis of connectivity (i.e., degree) since, by construction, in-degree and out-degree are identical for each vertex except the initial and final ones in an interval of seismic data analyzed. Accordingly, in-degree and out-degree are not distinguished from each other for the connectivity distribution. Thirdly, when small-worldness [2] is studied, directedness is ignored, tadpole-loops are removed and each multiple edge is replaced by a single edge. That is, the small-world picture is concerned with an undirected simple network reduced from a full earthquake network.

It has been reported in Refs. [1,3] that the earthquake networks thus constructed from the seismic data taken from California and Japan are scale free [4] and of the small-world type [2].

In this paper, we develop a discussion about relations between the exponents of the power-law connectivity distributions and spectral densities of the adjacency matrices of the reduced simple earthquake networks in California and Japan, in order to clarify the local structures of the networks. Carefully investigating the dependence of the connectivity exponent on the cell size, we show that the scaling relation derived by Dorogovtsev, Goltsev, Mendes and Samukhin [5], which is valid in the mean field approximation for locally treelike networks, holds well for the earthquake networks over ranges of the cell size. This result will be interpreted from the viewpoint of the hierarchical organization of the earthquake network [6].

We start our discussion with recalling the scaling relation presented in Ref. [5]. Let us consider an undirected simple scale-free network (i.e., without tadpole-loops and multiple edges) having $N$ vertices. In this case, the adjacency matrix, $A$, is symmetric and satisfies the following property: $(A)_{ij} = 1(0)$ if the $i$th and $j$th vertices are connected (unconnected). The connectivity of the $i$th vertex, $k_i$, is $k_i = \sum_{j=1}^{N} (A)_{ij}$. The connectivity distribution $P(k)$ is, in the large-$N$ continuous limit, given by $P(k) = (1/N) \sum_{i=1}^{N} \delta(k_i - k)$. On the other hand, the spectral density of the adjacency matrix reads $\rho(\lambda) = (1/N)\,\mathrm{tr}\,\delta(A - \lambda I)$, where $I$ denotes the $N \times N$ identity matrix. In the scale-free network, both of these quantities asymptotically obey power laws

$$P(k) \sim k^{-\gamma}, \qquad (1)$$

$$\rho(\lambda) \sim |\lambda|^{-\delta}, \qquad (2)$$

where $\gamma$ and $\delta$ are positive exponents. The authors of Ref. [5] have shown for a locally treelike network that, in the mean field approximation, holds $\rho(\lambda) = 2|\lambda| P(\lambda^2)$, leading to the scaling relation

$$2\gamma - \delta = 1. \tag{3}$$

In what follows, we wish to examine this relation for the reduced earthquake networks constructed from the data taken from California and Japan, which are currently available at http://www.data.scec.org/ and http://www.hinet.bosai.go.jp/. The space-time regions treated here are "between 1 January, 1984 and 31 December, 2006, 28.000N-39.414N latitude and 112.100W-123.624W longitude" and "between 3 June, 2002 and 15 August, 2007, 17.956N-49.305N latitude and 120.119E-156.047E longitude", respectively.

To observe if Eq. (3) holds for a real network, precise evaluation of the values of the exponents is essential. For an earthquake network, in particular, the dependencies of the exponents on the choice of the cell size should carefully be investigated. We have performed such an analysis by making use of the method of maximum likelihood estimation [7].

In Fig. 1, we present the connectivity distributions of the reduced simple earthquake networks. The scale-free nature as in Eq. (1) is observed. (Notice that the scale-free nature discussed in Ref. [1] is concerned with the full earthquake network containing tadpole-loops and multiple edges, and therefore it is different from the present one.)

The spectral density of the adjacency matrix is shown in Fig. 2. It is globally asymmetric but is almost symmetric around the origin. Its detailed behavior is depicted in Fig. 3. There, one can see that the spectral density decays as a power law as in Eq. (2). Slight asymmetry is reflected in the difference between two values of the exponent, $\delta_-$ and $\delta_+$. Also, the existence of the three peaks is noticed. The sharp peak at the zero mode, $\lambda = 0$, implies the presence of dead-end vertices (i.e., infinitely long "loops"), whereas the peaks around $\lambda = \pm 1$ correspond to long but finite loops. There might be other peaks, but they are not apparent in the figure.

Finally, the result of our analysis is summarized in Table 1. (The fact that the values of the cell size are taken different in the cases of California and Japan is due to the technical limitation on our computational power regarding the size of the adjacency matrices.) As can be seen, the scaling relation in Eq. (3) holds well over ranges of the cell size. This implies that the networks are locally treelike and the mean field approximation is good [5]. We emphasize that this result is consistent with the hierarchical organization of the earthquake network [6], which leads to the decay of the clustering coefficient with respect to the connectivity. This behavior of the clustering coefficient means that triangular loops tend not to be attached to hubs (i.e., the vertices of a main shock [1]) governing the gross property of the earthquake network. In fact, there is a striking empirical fact [1] that aftershocks associated with a main shock have a tendency to return to the locus of the main shock, geographically, without forming loops. Thus, the value of the clustering coefficient is suppressed around a hub, and accordingly the network becomes locally treelike.

To summarize, we have shown that the scaling relation presented in Ref. [5] holds well for the simple scale-free networks reduced from the full earthquake networks. This implies that the networks are locally treelike and the mean field approximation employed for deriving the scaling relation is well valid. We have interpreted this result in terms of the hierarchical organization of the earthquake networks. In the course of the present study, we have also performed careful analyses of the invariance of the exponent of the power-law connectivity distribution under change of the cell size as well as the scaling between the total number of vertices and the linear dimension of the cell size. These points are of relevance to the concept of renormalization and self-similarity of networks recently investigated in Ref. [8]. We wish to discuss this issue elsewhere [9].

*　　*　　*

The work of SA was supported in part by Grant-in-Aid for Scientific Research of the Japan Society for the Promotion of Science. NS acknowledges the support by Nihon University Individual Research Grant.

# Figure and Table Captions

Fig. 1 Log-log plots of the connectivity distributions of the simple earthquake networks. (a) California. The cell size is $15\text{ km}\times 15\text{ km}\times 15\text{ km}$. (b) Japan. The cell size is $50\text{ km}\times 50\text{ km}\times 50\text{ km}$. All quantities are dimensionless.

Fig. 2 The spectral densities of the adjacency matrices of the networks. (a) California. The cell size is $15\text{ km}\times 15\text{ km}\times 15\text{ km}$. (b) Japan. The cell size is $50\text{ km}\times 50\text{ km}\times 50\text{ km}$. Insets: the same spectral densities around the origins. All quantities are dimensionless.

Fig. 3 Log-log plots of the spectral densities in (a) California and (b) Japan corresponding to those in Fig. 2 in the regions of (-) negative $\lambda$ and (+) positive $\lambda$. All quantities are dimensionless.

Table 1 The exponents of the connectivity distributions and the spectral densities for some values of the cell size.

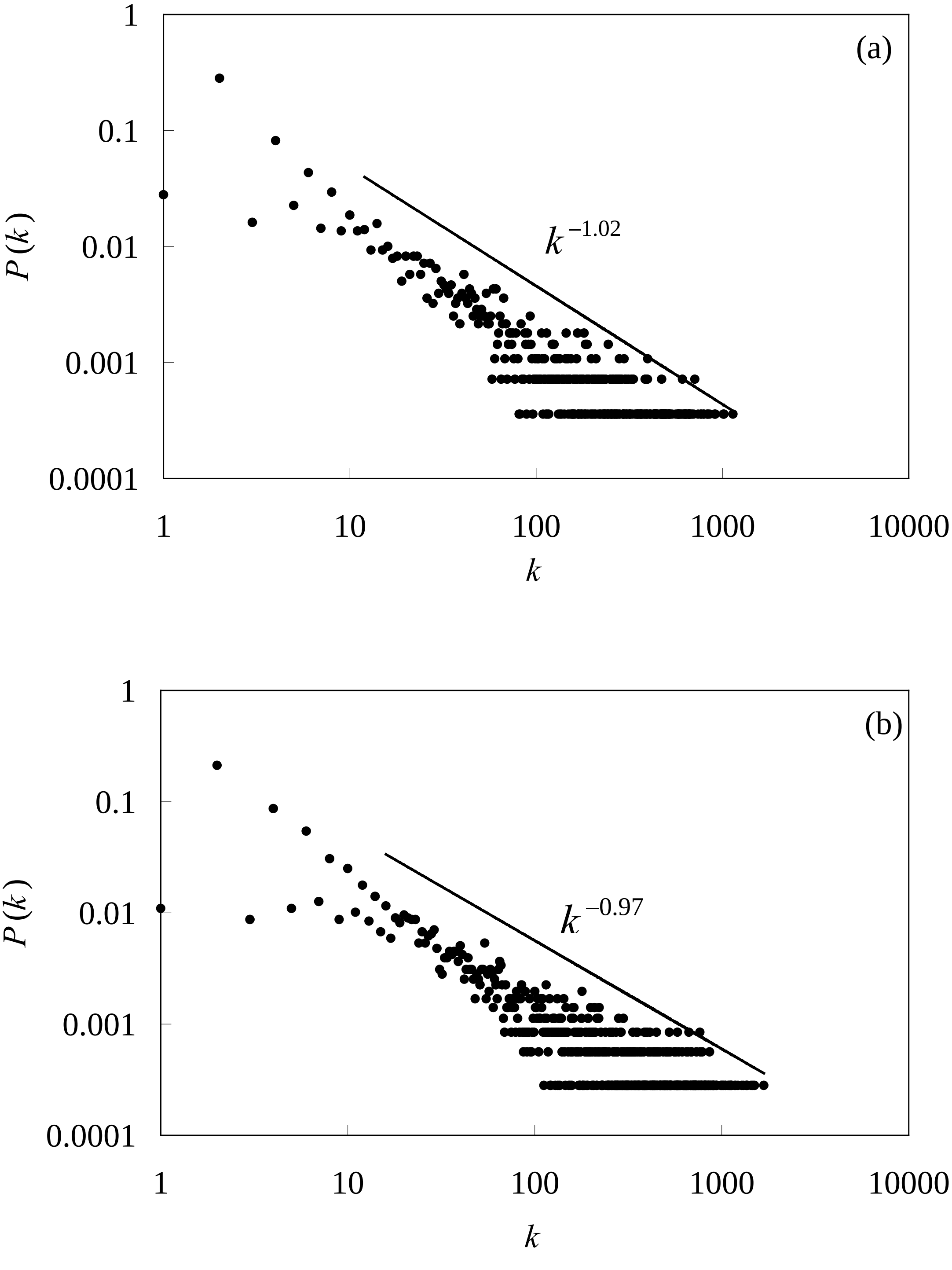

Fig. 1

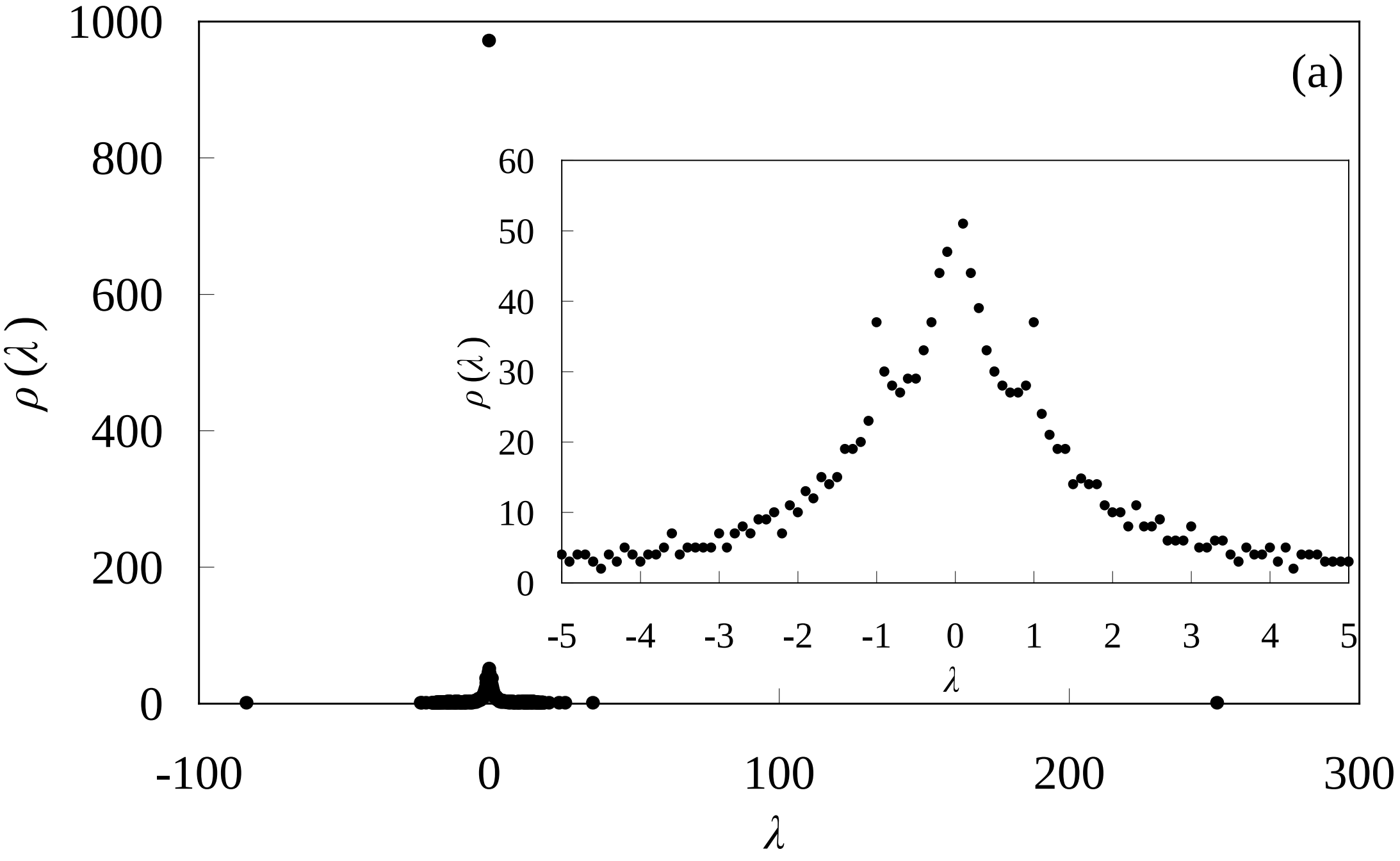

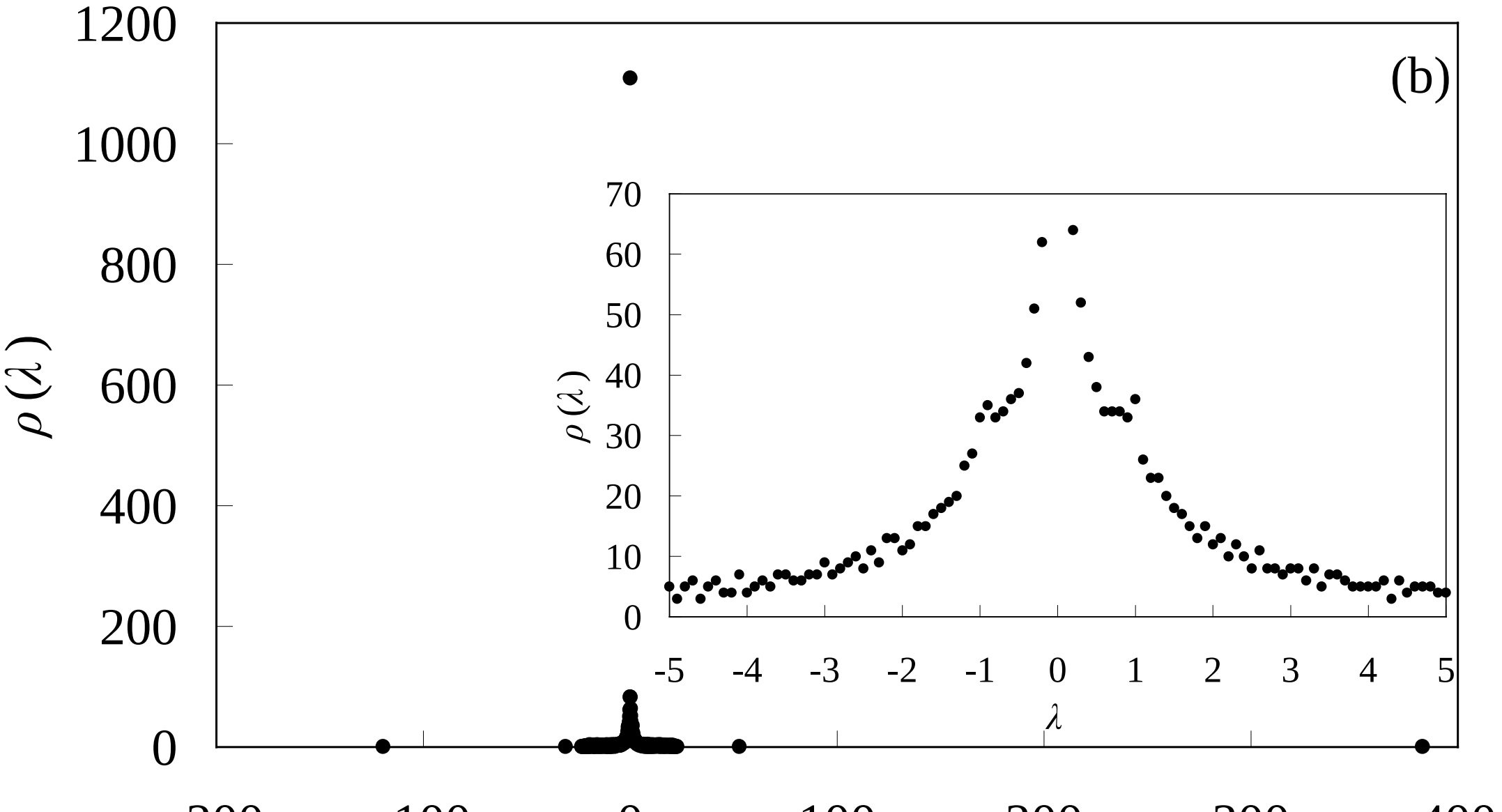

Fig. 2

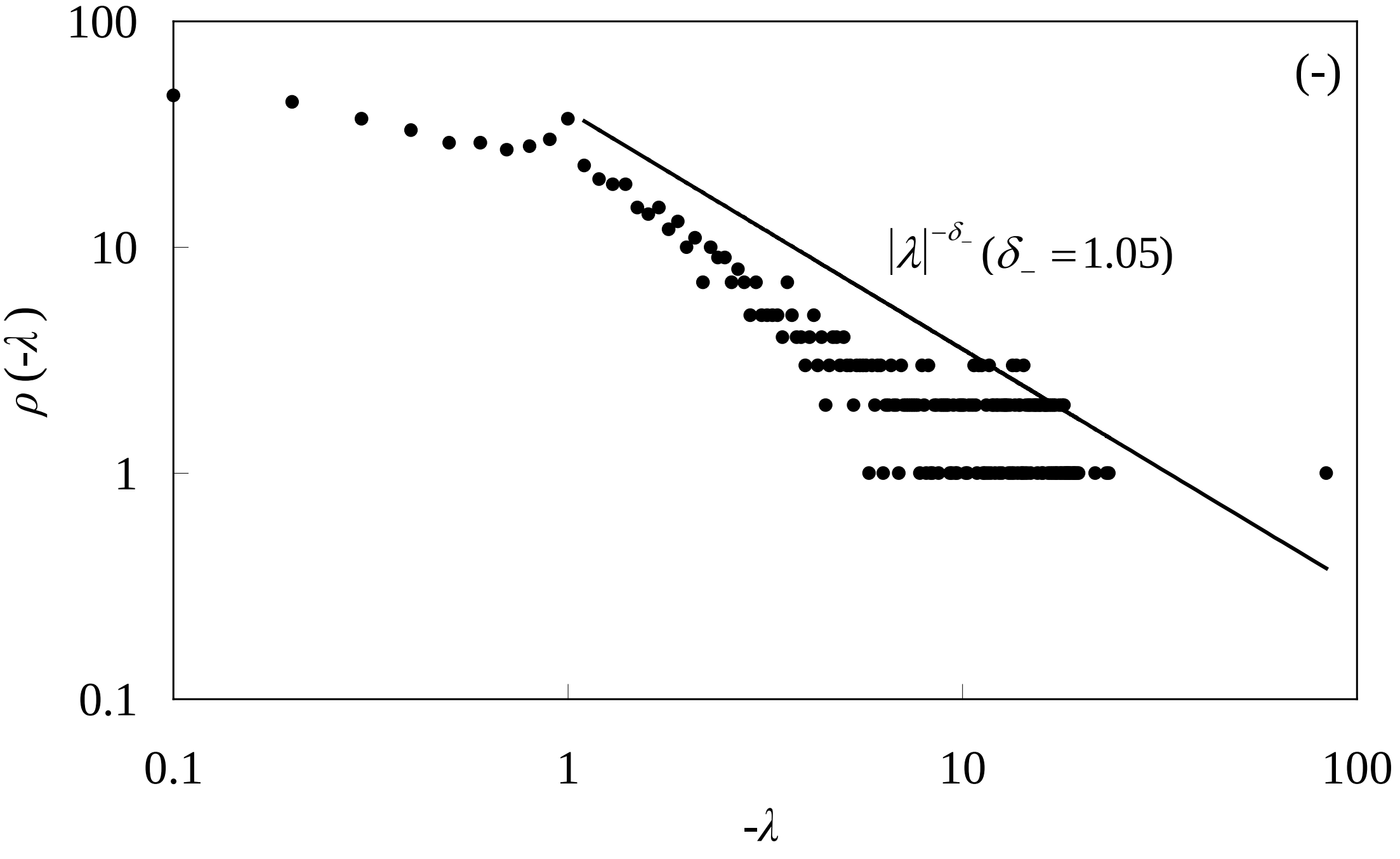

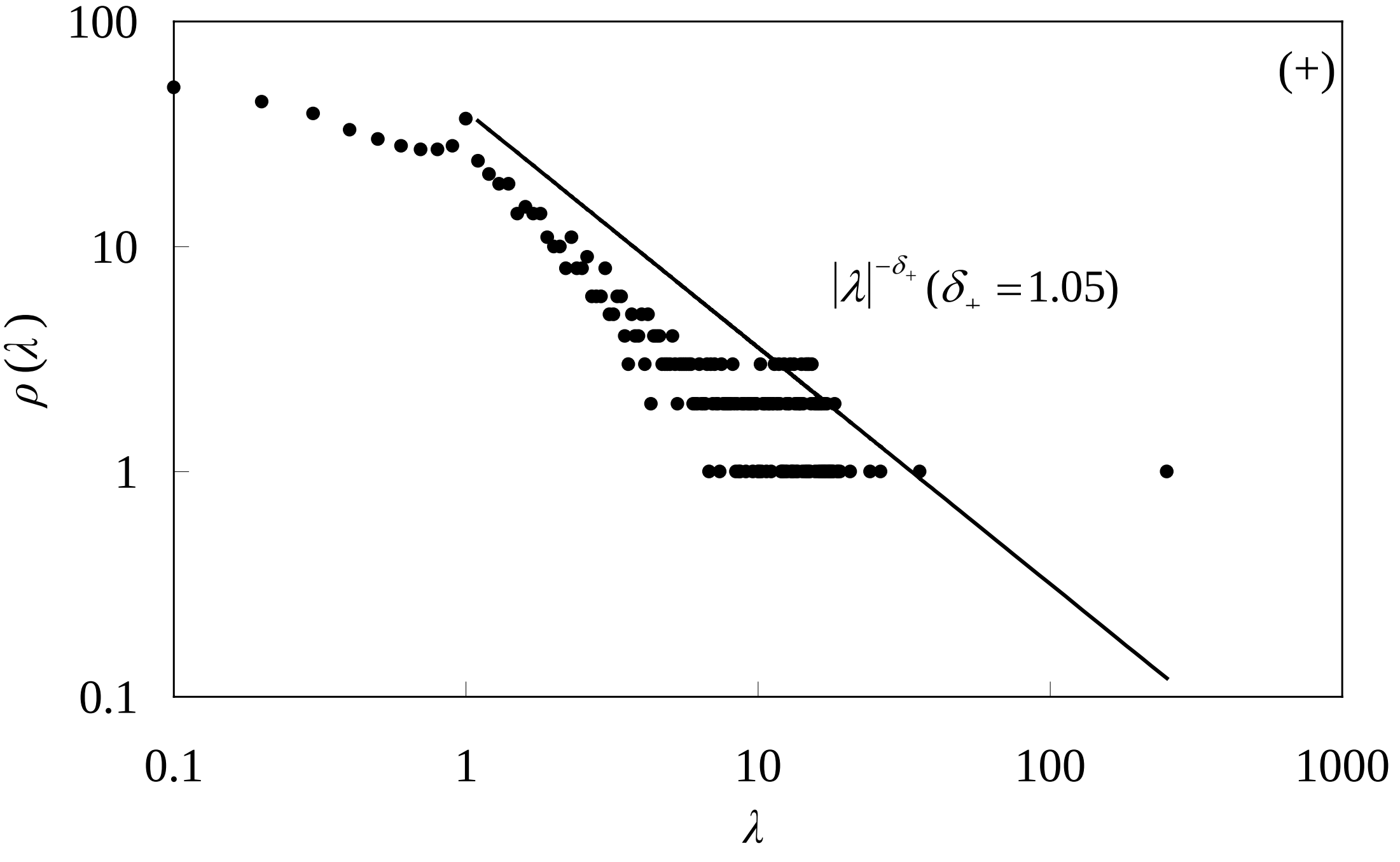

Fig. 3(a)

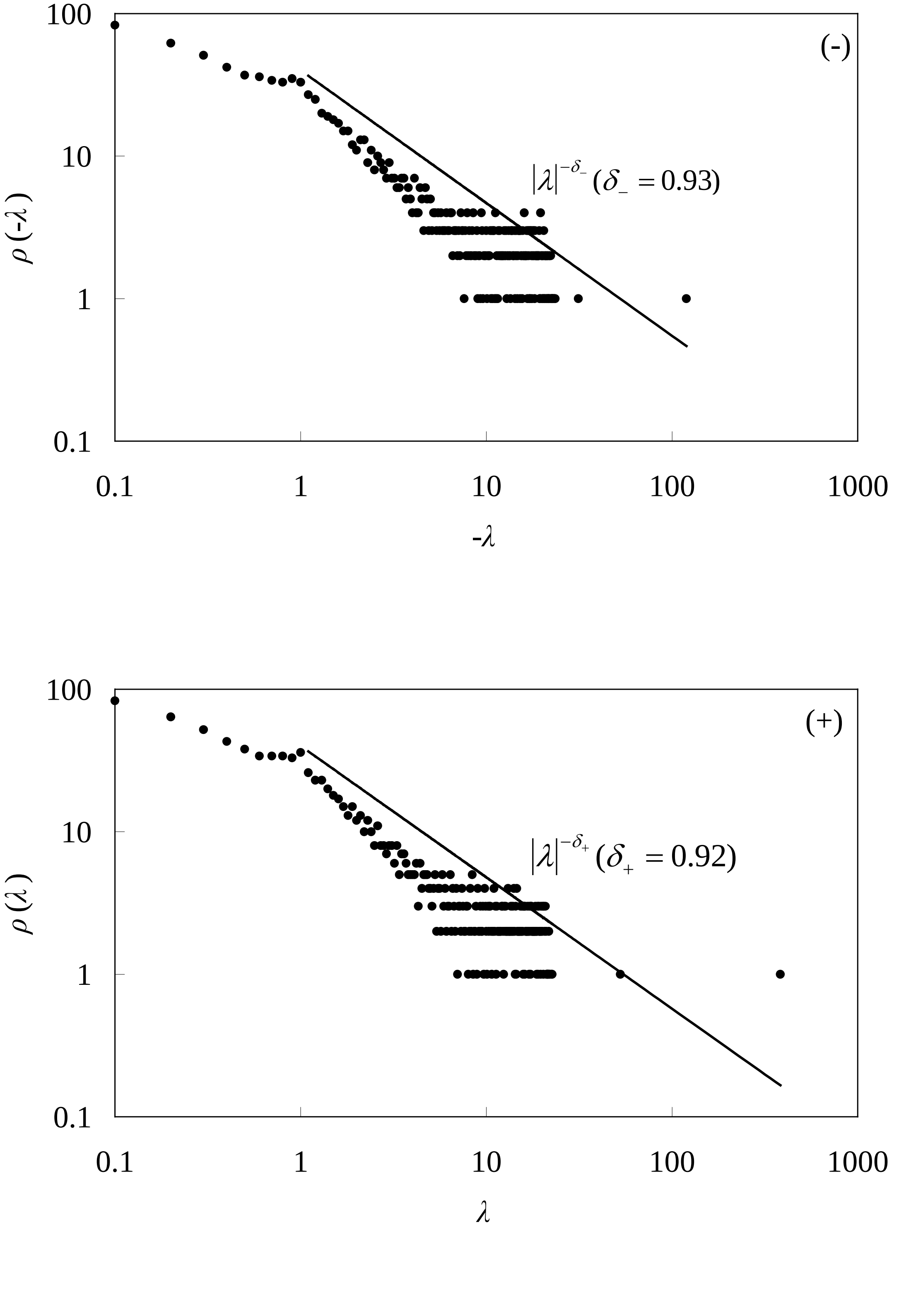

Fig. 3(b)

| | linear dimension of cell size (km) | $\delta_-$ | $\delta_+$ | $\gamma$ | $2\gamma-\delta_-$ | $2\gamma-\delta_+$ |
|---|---|---|---|---|---|---|
| California | 15 | 1.05±0.04 | 1.05±0.04 | 1.02±0.02 | 0.99±0.09 | 0.99±0.09 |
| | 20 | 0.82±0.05 | 0.83±0.05 | 0.90±0.02 | 0.97±0.12 | 0.97±0.12 |
| | 25 | 0.59±0.06 | 0.62±0.06 | 0.80±0.03 | 1.02±0.15 | 0.98±0.16 |
| Japan | 50 | 0.93±0.03 | 0.92±0.03 | 0.97±0.02 | 1.00±0.08 | 1.01±0.08 |
| | 60 | 0.78±0.04 | 0.79±0.04 | 0.88±0.02 | 0.98±0.09 | 0.98±0.09 |
| | 70 | 0.65±0.04 | 0.69±0.04 | 0.84±0.02 | 1.04±0.11 | 0.99±0.11 |

Table 1